\documentclass[paper]{ieice}
\pdfoutput=1
\usepackage{graphicx,xcolor}
\usepackage[fleqn]{amsmath}
\usepackage{newtxtext}
\usepackage[varg]{newtxmath}

\usepackage{comment}
\usepackage{url}

\field{D}
\title{A Software-based NVM Emulator Supporting Read/Write Asymmetric Latencies}
\authorlist{
 \authorentry[koshiba@namikilab.tuat.ac.jp]{Atsushi Koshiba}{n}{TUAT,AIST}
 \authorentry{Takahiro Hirofuchi}{n}{AIST}
 \authorentry{Ryousei Takano}{m}{AIST}
 \authorentry{Mitaro Namiki}{m}{TUAT}
}
\affiliate[TUAT]{The author is with Tokyo University of Agriculture and Technology, Tokyo, 184-8588 Japan.}
\affiliate[AIST]{The author is with National Institute of Advanced Industrial Science and Technology (AIST), Tsukuba, 305-8560 Japan.}

\begin{document}
\maketitle
\begin{summary}
Non-volatile memory (NVM) is a promising technology for low-energy and high-capacity main memory of computers.
The characteristics of NVM devices, however, tend to be fundamentally different from those of DRAM (i.e., the memory device currently used for main memory), because of differences in principles of memory cells.
Typically, the write latency of an NVM device such as PCM and ReRAM is much higher than its read latency.
The asymmetry in read/write latencies likely affects the performance of applications significantly.
For analyzing behavior of applications running on NVM-based main memory, most researchers use software-based emulation tools due to the limited number of commercial NVM products.
However, these existing emulation tools are too slow to emulate a large-scale, realistic workload or too simplistic to investigate the details of application behavior on NVM with asymmetric read/write latencies.
This paper therefore proposes a new NVM emulation mechanism that is not only light-weight but also aware of a read/write latency gap in NVM-based main memory.
We implemented the prototype of the proposed mechanism for the Intel CPU processors of the Haswell architecture.
We also evaluated its accuracy and performed case studies for practical benchmarks.
The results showed that our prototype accurately emulated write-latencies of NVM-based main memory: it emulated the NVM write latencies in a range from 200~ns to 1000~ns with negligible errors from 0.2\% to 1.1\%.
We confirmed that the use of our emulator enabled us to successfully estimate performance of practical workloads for NVM-based main memory, while an existing light-weight emulation model misestimated.
\end{summary}

\section{Introduction}

Recent trends of high-speed and many-core processors lead to an increasing demand for larger memory capacity.
Modern computer systems use DRAM for main memory while scaling up DRAM capacity is becoming difficult due to its refresh energy.
Because a DRAM cell holds its data as electric charge in a capacitor, periodically refreshing the cell is necessary to prevent data loss.
This energy overhead rapidly increases as DRAM scales up its capacity.
It is predicted that the refreshing energy occupies 50\% of the overall power consumption of a 64 GB DRAM module~\cite{Liu:2012:RRI:2337159.2337161}.
It is also reported that a server computer with 128 GB DRAM consumes more than 40\% of its energy consumption for its main memory~\cite{1250880}.
This energy-greedy characteristic of DRAM is an obstacle for future large capacity memory systems.

Non-Volatile Memory (NVM) is the key to overcome this energy constraint.
Some NVM devices with fast access latencies will have the potential to be used for the main memory of computers~\cite{ITRS}.
In addition, NVM does not require refreshing to keep its data, unlike DRAM.
This non-volatility prevents memory subsystems from wasting a large amount of energy.
Recent NVM technologies have attracted much attention not only in academia but also in the industry; new NVM products such as 3D-Xpoint are being developed~\cite{3dxpoint}.
For these reasons, NVM products are expected to achieve high-capacity and energy-efficient main memory systems.

\footnote[0]{\textbf{Note:} this paper extends our preliminary work published at NVMSA 2017~\cite{8064479}.
Specifically, we reimplemented a prototype of our emulator for the Intel Haswell processors, which previously targeted for an old processor architecture (i.e., Sandy Bridge) to verify the portability of our emulator for newer processor families.
Along with the reimplementation, we drastically improved the accuracy of the emulator by fixing bugs of cache miss measurement; the worst emulation error of the NVM write latency was mitigated from 28.6\% to 1.1\%.
Moreover, we conducted thorough experiments using various workloads.
All the parts of the paper are also thoroughly updated to improve the quality of the paper.
}

Although NVM is effective for energy reduction, current applications and system software, designed for DRAM-based main memory, will not efficiently work for future NVM-based main memory, due to its performance characteristics.
In particular, the gap between read latency and write latency is generally significant.
For example, phase change memory (PCM)~\cite{doi:10.1116/1.3301579} represents the state of a 1-bit cell (e.g., high or low) by changing its cell phase either of two phases: an amorphous phase (low) and a crystalline phase (high).
A read operation to a PCM cell just senses its resistance while a write operation applies an electrical pulse to the cell to heat it and change its phase.
Particularly, PCM recrystallization (changing from the amorphous phase to the crystalline phase) requires a long duration of pulsing.
Therefore, writing PCM typically requires much longer latency than reading.
The ITRS roadmap~\cite{ITRS} reports that the write latency of a typical PCM device is approximately 10x higher than its read latency. 
It also forecasts that writing PCM will be still 5x slower than reading it in 2026.
This gap possibly leads to performance degradation of write-intensive application programs.
For example, the results of our preliminary experiments (shown in Section \ref{sec:eval:speccpu2006} in this paper) showed that write-intensive workloads such as milc and libquantum experienced nearly 2x slower performance with NVM-based main memory in comparison to DRAM-based main memory.

To make use of future main memory with NVM, several researchers have tackled to find out new system software support and memory subsystems which are appropriate for NVM characteristics~\cite{7549421},~\cite{Dulloor:2014:SSP:2592798.2592814},~\cite{Condit:2009:BIT:1629575.1629589}.
However, no NVM-based main memories are commercially available.

Memory emulation tools are therefore essential for researchers to analyze/evaluate the performance of their proposals without actual NVM devices.
Although several simulation/emulation tools for NVM devices have been presented, these tools have problems for practical software research.
Cycle-accurate simulators~\cite{6296505},~\cite{7304374} are widely used among researchers.
While they can set read and write latencies independently in nanoseconds, these simulators are not appropriate for large-scale workloads because they are very time-consuming for system software emulation.
In contrast to heavy-weight simulators, Volos et al. proposed Quartz, which is a software emulator for NVM devices~\cite{Volos:2015:QLP:2814576.2814806}.
Quartz emulates NVM-based main memory using a computer with DRAM-based main memory.
It estimates the delays of the execution of a target process caused by accesses to an emulated NVM device and slows down the target process.
It uses the performance counters of a CPU processor to get information of memory accesses.
This emulation mechanism, slowing down a target process running on an operating system, is basically light-weight.
However, Quartz is unaware of the read/write latency gap of NVM devices.
Most CPU processors implement a write-back caching mechanism.
The CPU cores of a processor are not responsible for write-back to the main memory.
Instead, a cache controller handles it.
Quartz, using the performance counters of CPU cores, does not incorporate write-back information into the emulation model.

To overcome these shortcomings of existing emulation tools, this paper presents a light-weight NVM emulator that takes the read/write latency gap into account.
Unlike Quartz approach, our emulator classifies cache misses of a target process into two types: \textit{read-only} and \textit{write-back}.
The former performs only reading data from NVM, and the latter performs both reading and writing.
On NVM systems, write-back cache misses are expected to cause longer CPU stall cycles than the other.
To estimate the number of write-back cache misses, our emulator monitors not only CPU cache misses, but also the behavior of other components (prefetchers and cache controllers).
The emulator then calculates the additional delays caused by the two types of cache misses (read-only and write-back) respectively for the read/write latencies of an emulated NVM device.
This write-back aware emulation model enables an accurate emulation of NVM devices such as PCM.

To clarify the effectiveness of the proposed emulator, we developed a prototype of the proposed emulator on an Intel Xeon processor and conducted three experiments.
First, we evaluated the accuracy of the prototype.
We found that our prototype emulates the write latencies of NVM-based main memory in the range of 200 ns to 1000 ns with negligible errors of 0.1\% to 1.1\%.
Second, we applied our emulator to various workloads selected from SPECCPU 2006.
The results demonstrated that the use of our emulator successfully estimated the execution time of these workloads.
Third, we compared the proposed mechanism with Quartz using an in-memory database program, Memcached, as a case study of a realistic application.
In the experiment, we executed the original Quartz on our evaluation environment and applied it to Memcached.
We compared the evaluation results with our emulator.
We found that the use of Quartz misestimated the performance of Memcached running with NVM-based main memory.
These results show that our write-back aware mechanism has clear advantages in emulating NVM devices with asymmetric read/write latencies.

\section{Motivation}

In this section, we introduce the overview of memory access mechanisms and then explain how the gap of read/write latencies potentially impact on the performance of computers.

\subsection{Memory Access Mechanism}
\label{sec:motivation:memaccess}

  \begin{figure}[t]
    \begin{center}
      \includegraphics[width=84mm]{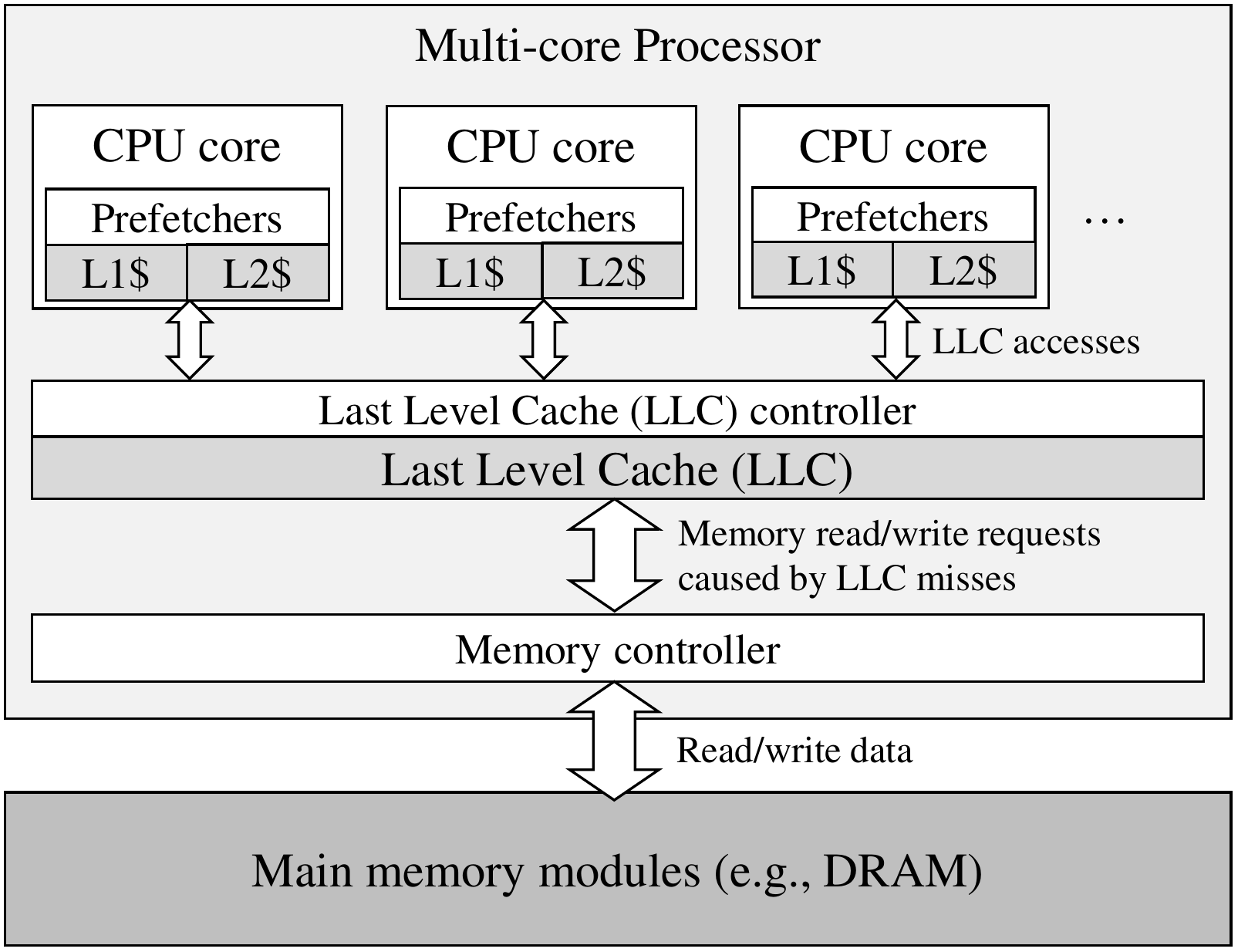}
    \end{center}
    \caption{Memory system structure of modern multi-core computer systems. Every CPU core of state-of-the-art processors may have more local cache levels (e.g., L3). We assume that NVM-based main memory is provided through memory modules and managed by the memory controller in the same manner as DRAM.}
    \label{fig:memory_structure}
  \end{figure}

We briefly explain a typical memory access mechanism in computers.
Fig. \ref{fig:memory_structure} shows the hardware structure of recent multi-core computer systems.
CPU cores are implemented in a multi-core processor and every CPU core has local caches (e.g., L1, L2).
Each CPU core has memory prefetchers, and it also executes instructuions in an out-of-order manner.
With these functions, two or more load/store instructions are sometimes performed concurrently and CPU cores avoid long stalls to access to the memory modules.
All CPU cores share the Last Level Cache (LLC), which is larger than local caches.
The LLC coherency among CPU cores is maintained by the LLC controller of the processor.
The LLC controller is also responsible for issuing read/write requests to the main memory when LLC misses occur.
The memory controller of the processor, receiving read/write requests from the LLC controller, operates main memory modules.
Note that CPU cores of recent processors have hardware performance counters, which measure the number of performance events (e.g., cache misses, stall cycles).
We assume that NVM-based main memory is byte-addressable in the same manner as DRAM-based main memory.
We also assume that both NVM and DRAM-based main memory modules are write-back cacheable; the caches in the processor hold modified data in cache lines and do not write the data to main memory modules until the cache lines are evicted.

In this memory architecture, memory references reaching to the main memory mostly occur when load/store instructions cause LLC misses.
A CPU core, executing a program, accesses memory data with load/store instructions.
When a CPU core executes a load or store instruction, it refers to a source or destination address of the main memory, which is specified by the instruction.
Because the data corresponding to the address may exist in caches, the CPU core first refers to its L1 cache.
If the data does not exist in the L1 cache, the CPU core refers to the next cache level (e.g., L2 and then LLC).
If the data does not exist even in the LLC, the CPU core triggers an LLC miss event.
It fetches a cache line of data (i.e., typically 64-bytes data) from the memory module.
When an LLC miss occurs, a cache controller selects an LLC line where new data should be maintained according to a certain cache management scheme (e.g., n-way set associative).
At the same time, the old data on the selected LLC line is evicted to make room for new data.

The procedure of an LLC miss differs depending on the state of the evicted LLC line.
If the state of the line is \textit{clean} or \textit{invalid}, the cache controller reads the new data from the memory module and overwrites it to the cache line.
On the other hand, if the state of the line is \textit{modified}, the controller not only reads new data from the module but also writes the modified line to the module in order to reflect the change to the main memory.
Therefore, we can find two types of LLC misses; one that just reads data from the memory module, and the other that induces a write-back.
We define the former and the latter as a \textit{read-only} LLC miss and a \textit{write-back} LLC miss, respectively.

\subsection{Impacts of Higher Write Latency}

  \begin{figure}[t]
    \begin{center}
      \includegraphics[width=84mm]{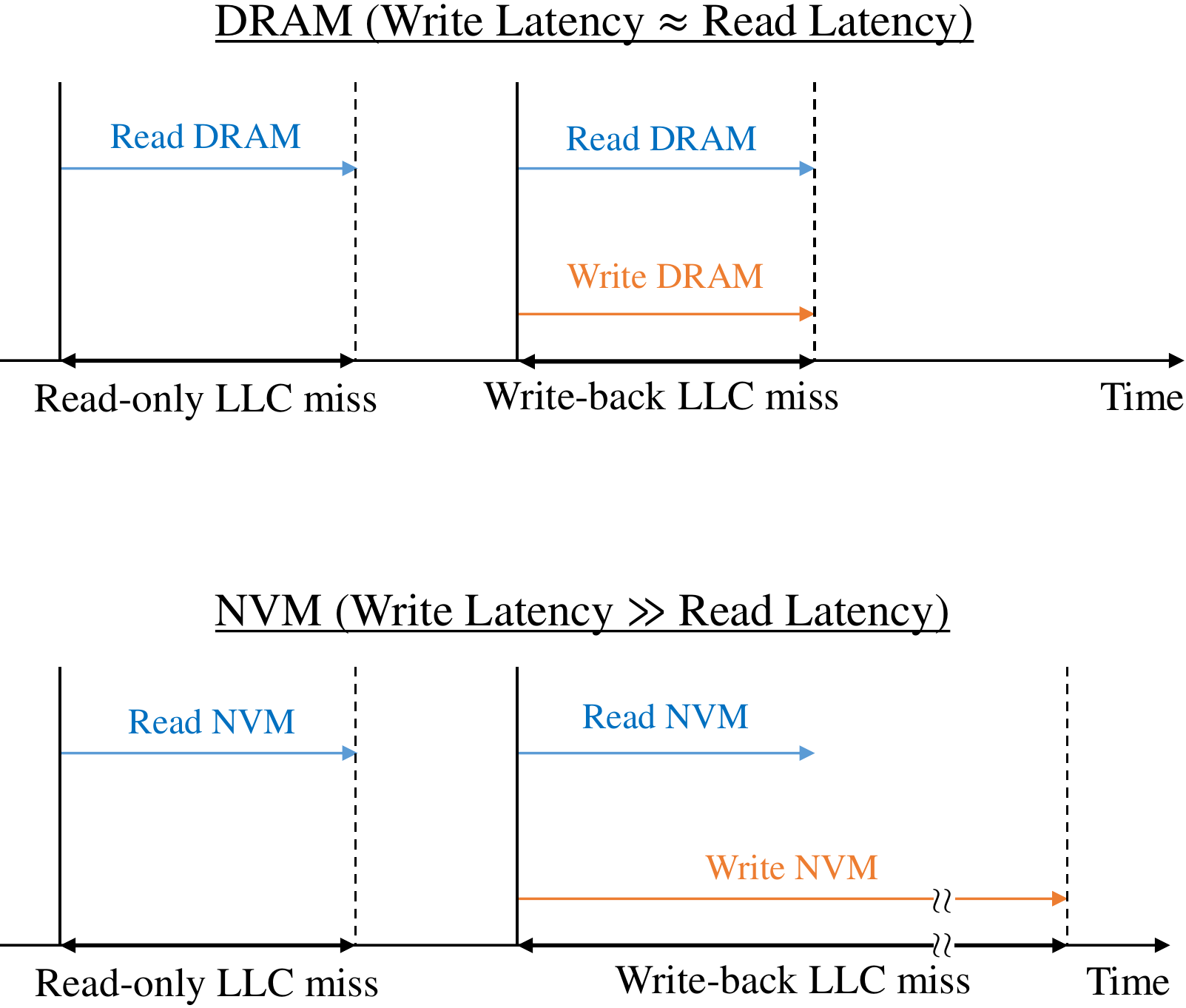}
    \end{center}
    \caption{The performance penalty upon an LLC miss in DRAM-based and NVM-based main memory systems, respectively.
    An NVM-based system likely experiences significant penalty upon a write-back LLC miss.}
    \label{fig:llc_miss}
  \end{figure}

The two types of LLC misses lead to the same latency with DRAM-based main memory because reading a new line and writing an old line are executed in parallel~\cite{Intel:optimization}.
Upon a write-back LLC miss, the LLC controller simultaneously starts reading a new line and writing an old line.
In DRAM-based main memory, the duration of a write-back LLC miss is the same as that of a read-only LLC miss, because read/write latencies of DRAM are the same.
However, if NVM devices such as PCM are used for main memory, the additional duration will be necessary upon a write-back LLC miss due to its higher write latency.
Fig.~\ref{fig:llc_miss} shows the difference of penalty time per one LLC miss between DRAM and NVM.
The upper part of Fig.~\ref{fig:llc_miss} shows the DRAM case where the write latency is almost the same as the read latency.
On the other hand, the lower part of Fig.~\ref{fig:llc_miss} shows the NVM case where the write latency is much longer than the read latency.
We assume that a write-back LLC miss in NVM-based main memory requires a longer period because the controller waits for the eviction of an old line.
Although the controller can temporarily hold write requests in a request queue to prevent write requests from interfering read requests, the queue will not work well for write-intensive applications because of its limited size.
Thus, if write-back LLC misses occur frequently, the CPU core that causes a write-back is forced to keep stalling until the old data eviction finishes.
This problem possibly influences the performance of application programs depending on their memory-access behavior.
For instance, our experimental results in Sec. \ref{sec:eval:speccpu2006} show that the execution time of libquantum, a write-intensive benchmark, becomes nearly 2x slower on NVM than on DRAM.

\subsection{Problem of Existing Work}
\label{sec:problem}
As described above, the read/write latency gap of NVM-based main memory possibly has a great impact on application performance. Analyzing its impact on performance is therefore indispensable for developing future NVM systems.
Because there are few numbers of commercial NVM products, researchers are forced to use emulation/simulation tools for their experiments.

However, there are some issues in existing tools to emulate the read/write latency gap.
The most common tool is cycle-accurate simulators.
These simulators are used with other CPU simulators and simulate full system behavior with NVM per CPU cycle~\cite{6218223},~\cite{6296505}.
This approach can set read/write latencies of main memory respectively, while it is too slow to emulate large-scale workloads.
For instance, we experienced that a simulation system using NVMain~\cite{6296505} with gem5~\cite{Binkert:2011:GS:2024716.2024718} took more than eight hours to finish a simulation of a tiny program, whose execution took only one second in reality.

  \begin{figure}[t]
    \begin{center}
      \includegraphics[width=84mm]{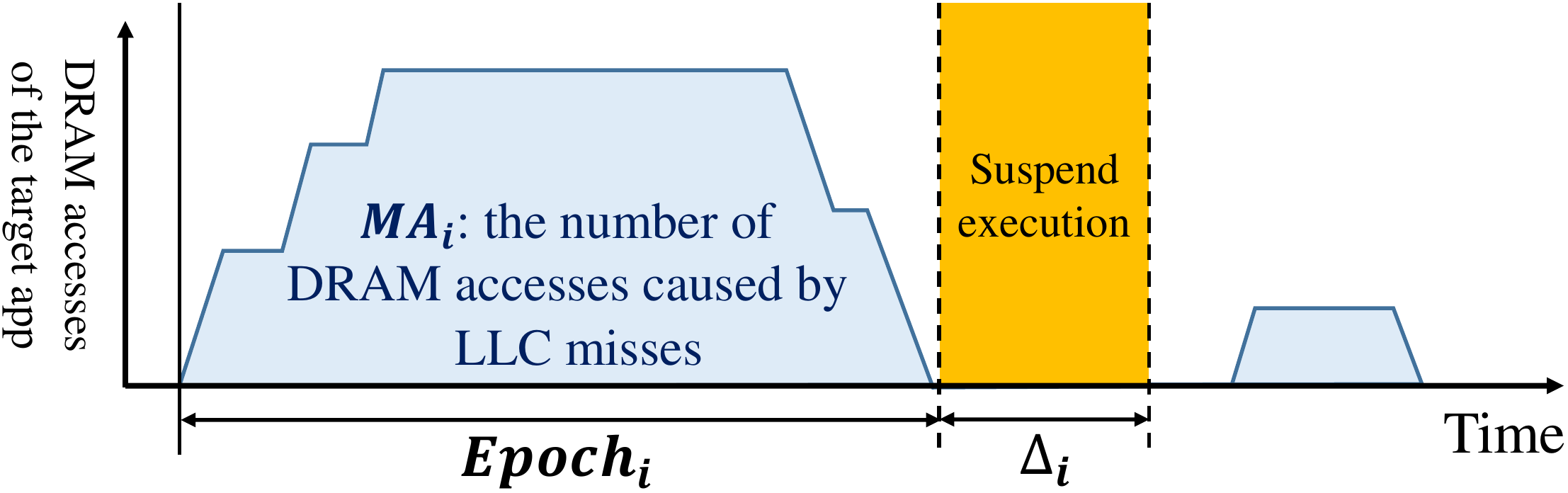}
    \end{center}
    \caption{The mechanism to delay the execution of a target process in Quartz.}
    \label{fig:quartz_approach}
  \end{figure}

On the other hand, Quartz~\cite{Volos:2015:QLP:2814576.2814806} is a light-weight emulation mechanism using hardware performance monitoring counters implemented in CPU cores of Intel processors.
To emulate a given NVM latency, Quartz inserts delays to the execution of a target process.
The inserted delays are based on the number of DRAM references obtained through performance counters of CPU cores.
Fig.~\ref{fig:quartz_approach} shows the Quartz emulation model.
Quartz measures the number of DRAM accesses caused by the target process using performance counters implemented in CPU cores at a specific interval named $Epoch$.
It then calculates the additional delay, $\Delta$, that is expected to be involved if the target process is executed with NVM-based main memory.
After the calculation, Quartz suspends the process execution until $\Delta$ elapses.
The overhead of this emulation mechanism is negligible for most use-cases.

The Quartz emulation model defines $\Delta_{i}$, the additional delay in $Epoch_{i}$, as Eq. (\ref{eq:delta}):
\begin{eqnarray}
\Delta_{i} = MA_{i} \times (NVM_{lat} - DRAM_{lat})
\label{eq:delta}
\end{eqnarray}
where $MA_{i}$ is the number of LLC misses during $Epoch_{i}$, which have caused CPU stalls of the CPU core executing the target process.
$NVM_{lat}$ and $DRAM_{lat}$ represents NVM access latency and DRAM access latency, respectively.
It should be noted that thanks to memory prefetching and out-of-order execution, an LLC miss does not necessarily involve a CPU stall. Thus, we need to count the number of the LLC misses involving CPU stalls, not the number of LLC misses.
To obtain $MA_{i}$, Quartz divides the number of the CPU stall cycles induced by LLC misses by DRAM access latency (in cycles):
\begin{eqnarray}
\begin{split}
MA_{i} = \frac{LLC\_STALL_{i}}{DRAM_{lat}}
\end{split}
\label{eq:ldm}
\end{eqnarray}
where $LLC\_STALL_{i}$ represents the total cycles of CPU core stalls caused by LLC misses.
The documentation of Intel CPUs~\cite{Intel:optimization} provides the equation to calculate $LLC\_STALL_{i}$ as follows:
\begin{eqnarray}
\begin{split}
LLC\_STALL_{i}& = L2_{stalls} \\
&\times \frac{W \times LLC_{miss}}{LLC_{hit} + W \times LLC_{miss}}
\end{split}
\label{eq:ldm_stall}
\end{eqnarray}
where $L2_{stalls}$ is the total number of core stall cycles caused by L2 cache misses, and $LLC_{hit}$ and $LLC_{miss}$ are the numbers of LLC hits and LLC misses of the core, and $W$ is the ratio of the LLC miss latency (DRAM access latency) to the LLC hit latency.

Although the Quartz approach has an advantage on the processing overhead over cycle-accurate simulators, it does not take the read/write latency gap into account.
The difficulty to support the latency gap stems from the lack of the capability in monitoring write-back activities through CPU cores;
CPU performance counters implemented in recent processors (e.g., Intel processors) do not support a performance event to measure write-back LLC misses of each CPU core.
The reason is considered that a modern processor assuming DRAM-based main memory does not need to pay attention to the write-back latency since DRAM write-back operations are completely hidden behind its read operations as shown in Fig.~\ref{fig:llc_miss}.
This fact makes it difficult for the emulation approach using CPU performance counters to analyze the impact of higher write latencies on the performance of a certain process.
To overcome this issue, we propose an emulation mechanism estimating per-core write-back LLC misses, which is not directly countable.

\section{Write-back Aware NVM Emulator}
This section proposes a light-weight emulation model that distinguishes write-back LLC misses and read-only LLC misses.

\subsection{Basic Idea}

We assume that write-back LLC misses lead to longer CPU stalls than read-only LLC misses.
To take the difference between read and write latencies into account, our emulation model monitors two types of LLC misses respectively unlike the Quartz emulation model.
Our model allows users to evaluate applications performance with NVM devices whose read/write access latencies are asymmetric.

  \begin{figure}[t]
    \begin{center}
      \includegraphics[width=84mm]{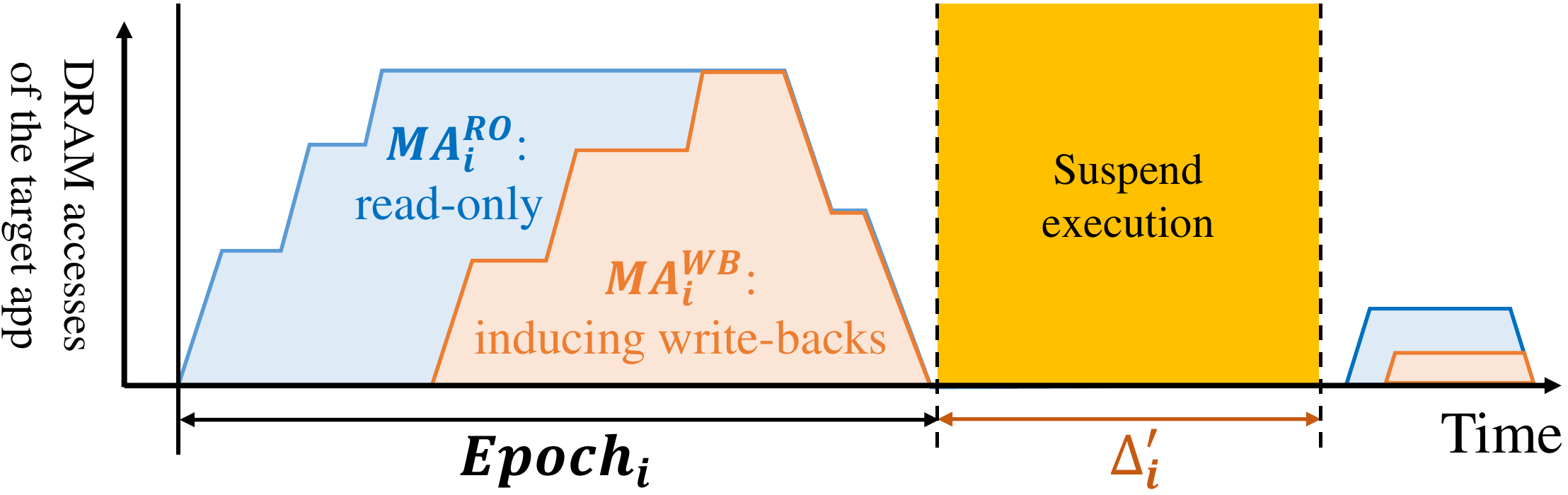}
    \end{center}
    \caption{The mechanism to delay the execution of a target process in the proposed emulation model. It distinguishes LLC misses into two types: read-only and write-back. The latter, LLC misses inducing write-backs, are expected to cause longer CPU stalls than the former. }
    \label{fig:proposed_approach}
  \end{figure}

Our emulator injects delays into a target process depending on the number of LLC misses in the same manner as Quartz.
However, unlike Quartz, our model divides LLC misses into two types; one just reads data from memory modules (read-only) and the other induces both reading and writing (write-back) as shown in Fig.~\ref{fig:proposed_approach}.
$MA_{i}^{WB}$ in Fig.~\ref{fig:proposed_approach} is the number of write-back LLC misses, and $MA_{i}^{RO}$ is the number of read-only LLC misses within $Epoch_{i}$.
Note that $MA_{i}^{WB}$ and $MA_{i}^{RO}$ represents the number of LLC misses that actually cause CPU stalls.
These two types of LLC misses satisfy the following condition:
\begin{eqnarray}
MA_{i} = MA_{i}^{WB} + MA_{i}^{RO}
\label{eq:mem_accesses}
\end{eqnarray}
We assume that the write-back LLC misses make CPU cores stalled for a longer period than the read-only LLC misses.

Let $NVM^{Write}_{lat}$ be the average NVM write latency and let $NVM^{Read}_{lat}$ be the average NVM read latency ($NVM^{Write}_{lat} \gg NVM^{Read}_{lat}$), our model represents the additional delay $\Delta_{i}^{'}$ as follows:
\begin{eqnarray}
\begin{split}
\Delta_{i}^{'} &= MA_{i}^{WB} \times (NVM_{lat}^{Write} - DRAM_{lat}) \\
&+ MA_{i}^{RO} \times (NVM_{lat}^{Read} - DRAM_{lat})
\end{split}
\label{eq:mydelta}
\end{eqnarray}
To calculate the value of $\Delta_{i}^{'}$,
the emulator needs to periodically estimate $MA_{i}^{WB}$ and $MA_{i}^{RO}$ of the target process at run-time.
However, performance counters of CPU cores cannot measure the number of write-back LLC misses because of the cache architecture.
Therefore, we present a way to estimate the number of write-back LLC misses and achieve a write-back aware NVM emulator.

\subsection{Run-time Estimation of Read-only/Write-back Memory Accesses}
This section describes how to calculate the two types of LLC misses ($MA_{i}^{RO}$ and $MA_{i}^{WB}$) respectively.
Our emulation model enables the calculation by making use of performance counters of the LLC controller in addition to information obtained from performance counters of CPU cores.
Our model defines $MA_{i}^{WB}$ and $MA_{i}^{RO}$ as shown in Eq. (\ref{eq:ldms}):
\begin{eqnarray}
\begin{split}
&MA_{i}^{WB} = \frac{LLC\_STALL_{i}^{WB}}{DRAM_{lat}}, \\
&MA_{i}^{RO} = \frac{LLC\_STALL_{i}^{RO}}{DRAM_{lat}}
\end{split}
\label{eq:ldms}
\end{eqnarray}
where $LLC\_STALL_{i}^{WB}$ and $LLC\_STALL_{i}^{RO}$ are the total cycles of CPU core stalls caused by write-back LLC misses and read-only LLC misses, respectively.
To calculate $LLC\_STALL_{i}^{WB}$ and $LLC\_STALL_{i}^{RO}$, our model extends Eq. (\ref{eq:ldm_stall}).
$LLC_{miss}$ in Eq. (\ref{eq:ldm_stall}) can be classified into two types (write-back and read-only) as we have already described in Sec. \ref{sec:motivation:memaccess}.
Our model then defines $LLC\_STALL_{i}^{WB}$ and $LLC\_STALL_{i}^{RO}$ as Eq. (\ref{eq:ldm_stall_wb}) and Eq. (\ref{eq:ldm_stall_ro}):
\begin{eqnarray}
\begin{split}
\hspace{-3mm} LLC\_STALL_{i}^{WB}& = L2_{stalls} \\
&\times \frac{W \times LLC_{miss}^{WB}}{LLC_{hit} + W \times LLC_{miss}}
\end{split}
\label{eq:ldm_stall_wb}
\end{eqnarray}
\begin{eqnarray}
\begin{split}
\hspace{-3mm} LLC\_STALL_{i}^{RO}& = L2_{stalls} \\
&\times \frac{W \times (LLC_{miss} - LLC_{miss}^{WB})}{LLC_{hit} + W \times LLC_{miss}}
\end{split}
\label{eq:ldm_stall_ro}
\end{eqnarray}
where $LLC_{miss}^{WB}$ is the total number of write-back LLC misses.

Due to the lack of performance monitoring events of CPU cores, $LLC_{miss}^{WB}$ cannot be counted directly.
Therefore, our model estimates $LLC_{miss}^{WB}$ using other available monitoring functions.
To estimate $LLC_{miss}^{WB}$, there are two key factors: (1) the number of write-backs within a certain period, and (2) the degree of contribution of the target process to these write-backs.
To measure the factor (1), our model uses an \textit{uncore} performance counter implemented on the cache controller.
Intel processors such as Intel Xeon have LLC controllers called LLC coherency engines (CBo)~\cite{E5-v3:Uncore}.
Because CBo counters monitor the number of cache lines written back to the memory modules, they enable our model to measure the factor (1) directly.
Next, to estimate the factor (2), our model measures the number of all LLC misses caused by CPU cores and their prefetchers in the system.
We expect that the degree of contribution of a certain CPU core to write-backs can be estimated based on the proportion of its LLC misses to the whole.
Assuming that a certain core causes 40,000 LLC misses in an epoch and the total number of LLC misses in the same epoch is 200,000, the number of LLC misses caused by the core occupies 20\% of all the LLC misses.
Since write-back requests are induced by LLC misses, the number of write-backs caused by the core in this epoch is expected to be 20\% of all the write-backs.
Thus, if the total number of write-backs in this epoch is 50,000, the number of write-back LLC misses of the core is expected to be 10,000.
Based on these considerations, our model estimates $LLC_{miss}^{WB}$ with Eq. (\ref{eq:LLCmiss_wb}):
\begin{eqnarray}
\begin{split}
\hspace{-3mm} LLC_{miss}^{WB}& = WB \\
&\times \frac{LLC_{miss}}{\sum_{i=0}^{n-1}LLC_{miss,cpu_{i}} + \sum_{i=0}^{n-1}LLC_{miss,PF_{i}}}
\end{split}
\label{eq:LLCmiss_wb}
\end{eqnarray}
where $WB$ is the total number of write-back operations by the cache controller, $n$ is the number of CPU cores of a processor, $\sum_{i=0}^{n-1}LLC_{miss,cpu_{i}}$ is the sum of the numbers of LLC misses caused by every CPU core, $\sum_{i=0}^{n-1}LLC_{miss,PF_{i}}$ is the sum of the numbers of LLC misses caused by every prefetcher.
Eq. (\ref{eq:LLCmiss_wb}) calculates the ratio of LLC misses of the target process to LLC misses of the whole system and then multiplies the ratio and the number of write-backs.
Thus, the equation gives us the estimated number of write-back LLC misses caused by a specific process.

\subsection{Applying to an Intel Processor}

\begin{table}[t]
  \caption{The performance monitoring events of the Haswell architecture family used in the proposed emulator.}
  \begin{center}
    \begin{tabular}{l|l} \hline \hline
      \multicolumn{2}{c}{Performance events of CPU counters~\cite{Intel:manual}} \\ \hline \hline
      $L2_{stalls}$   & CYCLE\_ACTIVITY:STALLS\_L2\_PENDING \\ \hline
      $LLC_{hit}$     & MEM\_LOAD\_UOPS\_L3\_HIT\_RETIRED: \\
                      & XSNP\_NONE \\ \hline
      $LLC_{miss}$,   & MEM\_LOAD\_UOPS\_L3\_MISS\_RETIRED: \\
      $LLC_{miss,cpu_{i}}$ & LOCAL\_MEM \\ \hline
      $LLC_{miss,PF_{i}} + $  & OFFCORE\_RESPONSE\_0  \\
      $LLC_{miss,cpu_{i}}$   & (offcore rsp: 0x3FB84003F7) \\ \hline \hline
      \multicolumn{2}{c}{Performance events of CBo (LLC controller) counters~\cite{E5-v3:Uncore}} \\ \hline \hline
      $WB$    & LLC\_VICTIMS.M\_STATE \\ \hline
    \end{tabular}
  \end{center}
  \label{table:pmc_param}
\end{table}

We implemented a prototype of our emulator for the Intel Haswell architecture.
Table~\ref{table:pmc_param} shows the performance counter events corresponding with the variables of the above equations~\cite{Intel:manual},~\cite{E5-v3:Uncore}.
$DRAM_{lat}$ and $W$ are static values relying on the performance of a given machine and can be measured using a tool such as Intel Memory Latency Checker (MLC)~\cite{Intel:mlc}.

  \begin{figure}[t]
    \begin{center}
      \includegraphics[width=84mm]{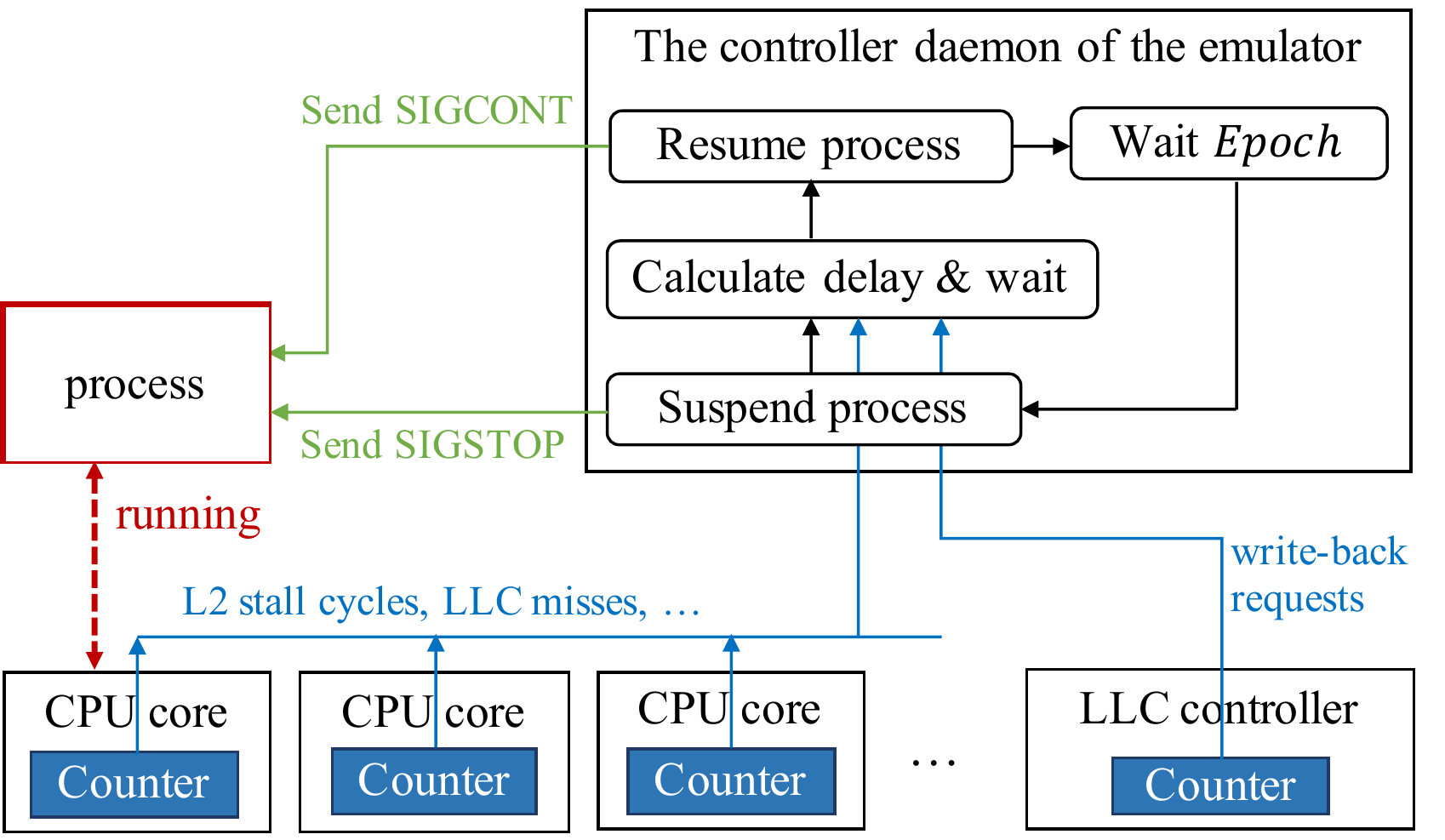}
    \end{center}
    \caption{The overview of the emulation mechanism of our emulator.}
    \label{fig:implementation}
  \end{figure}

Fig. \ref{fig:implementation} shows the execution flow of the controller daemon of the emulator.
Both the controller daemon and the emulated process are running on the same multi-core processor during the emulation.
The controller daemon periodically calculates and injects an additional delay at every fixed interval ($Epoch$).
When $Epoch$ elapses, the controller daemon suspends the execution of the target process.
It then reads performance counters of CPU cores and the LLC controller to obtain values of performance events shown in Table \ref{table:pmc_param}.
It calculates the additional delay from the obtained values using our emulation model.
The target process is suspended until its idle time reaches the calculated delay.
Finally, the controller daemon resumes the target process and waits for the next $Epoch$.
The controller daemon uses POSIX signals to suspend the process execution and to resume it.
Our prototype is portable because other Intel processor families are equipped with performance counters that support the equivalent events.

\section{Evaluation}
To verify the effectiveness of our emulation model, we evaluated the prototype of the proposed emulator using a computer with an Intel Xeon E2637 v3 processor.
The processor is the Intel Haswell architecture.
Table~\ref{table:env} shows the detail of our evaluation environment.
We used Intel MLC to measure the values of $DRAM_{lat}$ and $W$.
In the experiments, we configured the $Epoch$ parameter so that our emulator can make a good balance between accuracy and calculation overhead.
Because setting a longer Epoch value will increase the possibility that the emulator fails to track short temporal changes of memory access behavior of a workload, a shorter Epoch value is preferable in this sense.
Although, a shorter Epoch value results in the increase of CPU load due to the calculation overhead.
We observed that in the experiments 20 ms of Epoch was appropriate to accurately emulate NVM with negligible calculation overhead.
In other situations, it may be necessary to tune up an Epoch value to obtain sufficient accuracy, especially for workloads whose memory access behaviors frequently change.
We will conduct further investigation on the relationship between the emulation accuracy and Epoch in future work.

We used the machine exclusively for the emulation and also configured the operating system to stop unnecessary services.
We consider that the cache hit ratio in the experiments will be close to that of a machine with a real NVM device.
We focus this paper on the evaluation of our emulator for single-threaded (or single-process) workloads.
However, if the emulator is applied to multi-threaded (or multi-processes) workloads, the temporal suspension of each thread may change the behavior of cache contention.
This may result in the difference in the cache hit ratio and degrade the accuracy of emulation.
We will report the feasibility of the emulator for such applications in our upcoming work.

\begin{table}[t]
  \caption{Our Evaluation Environment}
  \begin{center}
    \begin{tabular}{l|l} \hline \hline
      Processor  & Intel Xeon E5-2637 v3\\ \hline
      OS     & CentOS 6.10 (Linux 2.6.32)\\ \hline
      Memory & 32GB DDR4 RAM @2400MHz \\ \hline
      $Epoch$   & 20 ms \\ \hline
      $DRAM_{lat}$& 121.7 ns\\ \hline
      $W$     & 4.14\\ \hline \hline
    \end{tabular}
  \end{center}
  \label{table:env}
\end{table}

\subsection{A Tool to Measure Write-back Latency}
  \begin{figure}[t]
    \begin{center}
      \includegraphics[width=84mm]{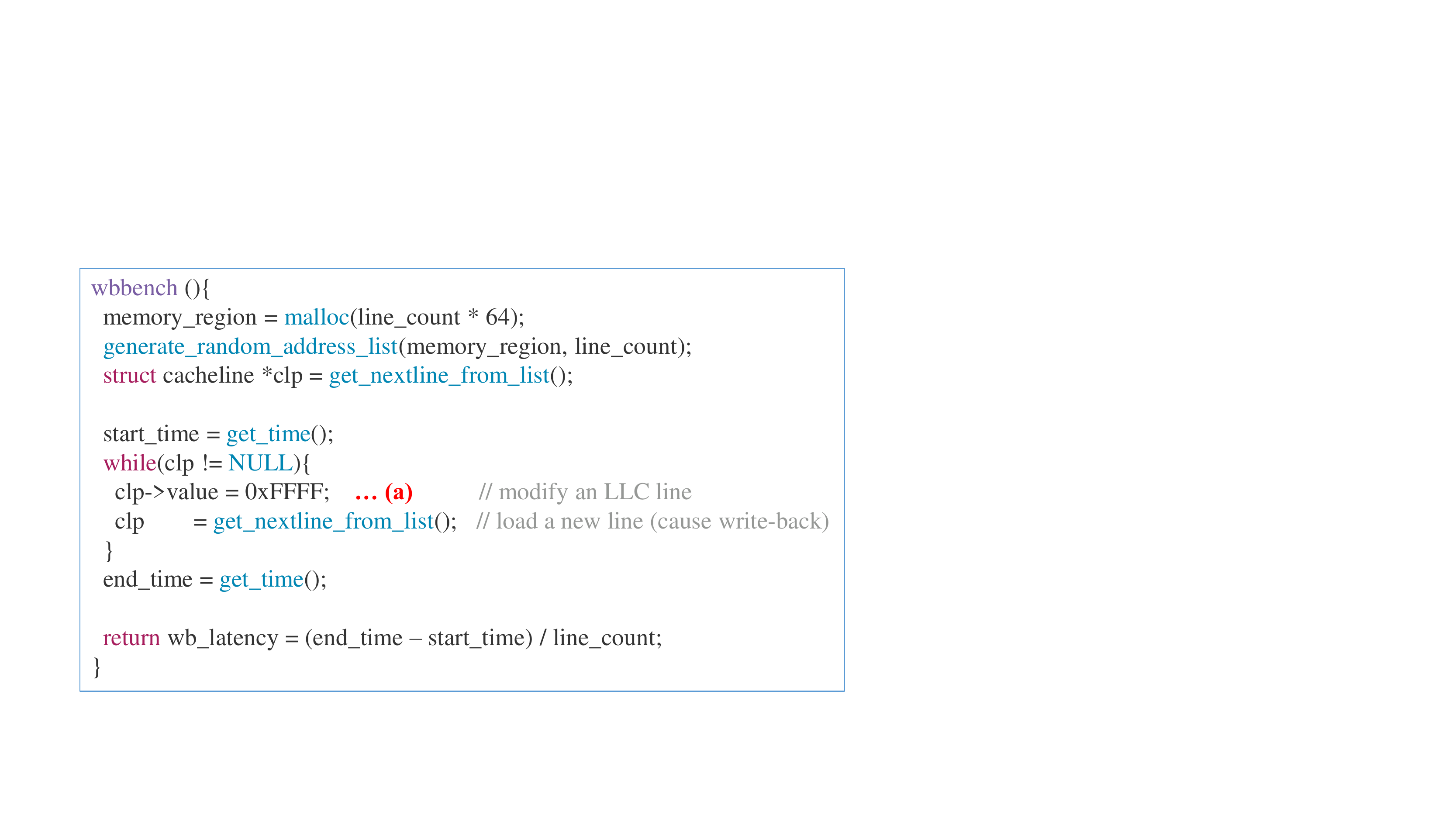}
    \end{center}
    \caption{Pseudo code of wbbench.}
    \label{fig:wbbench}
  \end{figure}

To evaluate the precision of our model emulating the NVM write latency, we developed a tool named \textit{wbbench} that measures the average latency of write-back LLC misses.
Fig.~\ref{fig:wbbench} shows the pseudo code of wbbench.
In order to accurately measure cache miss latencies, wbbench is carefully designed to suppress the effect of prefetching and out-of-order execution.
First, wbbench calls malloc() to reserve a certain amount of memory region.
It then calls generate\_random\_address\_list() to split the memory region into a linked list of cache-line aligned objects (i.e., struct cacheline).
Each cache line object is aligned to the size of an LLC line (64 bytes).
The cacheline objects of the linked list are arranged in a random order; a cacheline object points to the next one likely located at a distant address.
While executing the while() loop, wbbench writes a value to the cache line object that is currently referred by a pointer (clp).
Next, it calls get\_nextline\_from\_list() to refer to the address of the next cache line object in the list and store it in the pointer, which causes an LLC miss with a line eviction.
Since the cache line objects in the list are arranged randomly, wbbench suppresses the effect of memory prefetching and out-of-order execution.
The memory access prefetching is not effective for random access to cache lines.
The out-of-order execution of CPU does not work effectively for the pointer traversal of a link list.
The size of the memory region is set to be sufficiently larger than the size of LLC; at each iteration of the while() block, a write-back LLC miss occurs at a high probability.
Wbbench measures the total elapsed time during the while() loop and calculates the average write-back latency.

We also implemented robench, a tool to measure the average latency of read-only LLC misses.
Robench code is almost the same as the wbbench code.
The only difference is that robench does not modify a cache line (i.e., skips the line marked as (a) in the pseudo code).
Since LLC misses caused by robench do not induce write-backs, robench can measure the read-only LLC miss latency.

To confirm the accuracy of latency measurement of wbbench and robench, we measured the LLC miss latencies of a computer with DRAM-based main memory.  The read-only and write-back latencies should be the same in a DRAM reference.
For comparison, Intel MLC was also used to measure \textit{a DRAM latency}, which does not distinguish read/write latencies.
Table \ref{table:dram_latency} shows DRAM access latencies measured on our experimental environment.
Measurement errors of wbbench and robench in comparison with Intel MLC is 1.5 ns (1.2\%) and 0.9 ns (0.7\%), respectively.
We observed that their output results are very close to those of the Intel's proprietary measurement program.
The results indicate that our latency measurement programs are sufficiently accurate.

\begin{table}[t]
  \caption{DRAM access latencies measured with different tools. *Intel MLC does not distinguish the read/write latencies.}
  \begin{center}
    \begin{tabular}{c|c|c} \hline \hline
     & Intel MLC & Our tools \\ \hline \hline
    Measured read latency & 121.7 ns & 122.6 ns (with robench) \\ \hline
    Measured write latency & * & 123.2 ns (with wbbench) \\ \hline \hline
    \end{tabular}
  \end{center}
	\label{table:dram_latency}
\end{table}

\subsection{Validating Accuracy of Emulation}
\label{sec:eval:accuracy}
We evaluated the accuracy of the proposed emulation mechanism using wbbench and robench.
We set target read/write latencies of the emulator and then measured actually-emulated latencies by wbbehcn and robench, i.e., wbbench or robench were executed in our latency emulator.
If our prototype can accurately emulate the write latency of NVM, a target write latency and its actually-emulated latency become very close.
To ensure that every get\_nextline\_from\_list() call induces an LLC miss, we set the size of the memory region reserved by wbbench/robench to 30 MB, which is twice as large as the LLC size of our environment (15 MB).

Table~\ref{table:wbaware_latency} shows the evaluation results.
The emulated write latency was changed from 200 ns to 1000 ns while the read latency is the same as the actual DRAM latency.
When applying our emulator to wbbench, the NVM write latencies were emulated with errors of 0.1\% to 1.1\%.
In addition, when applying our emulator to robench, the NVM read latencies were emulated with errors of 2.7\% to 5.4\%.
These results show that our mechanism can emulate asymmetric read/write latencies with negligible errors.

\begin{table}[t]
  \caption{NVM latencies configured by our prototype and measured with wbbench/robench. }
  \begin{center}
    \begin{tabular}{c||c|c|c|c} \hline \hline
      Configured & \multicolumn{2}{c|}{wbbench} & \multicolumn{2}{c}{robench} \\ \cline{2-5}
      read/write lat. & Measured lat. & error & Measured lat. & error \\ \hline \hline
      122 ns/200 ns           & 202.1 ns & 1.1  \% & 125.3 ns & 2.7 \% \\ \hline
      122 ns/300 ns           & 300.4 ns & 0.1  \% & 125.6 ns & 3.0 \% \\ \hline
      122 ns/400 ns           & 399.2 ns & -0.2 \% & 125.8 ns & 3.1 \% \\ \hline
      122 ns/500 ns           & 497.8 ns & -0.4 \% & 126.4 ns & 3.6 \% \\ \hline
      122 ns/1000 ns           & 988.7 ns & -1.1 \% & 128.6 ns & 5.4 \% \\ \hline \hline
    \end{tabular}
  \end{center}
  \label{table:wbaware_latency}
\end{table}

  \begin{figure*}[t]
    \begin{center}
      \includegraphics[width=170mm]{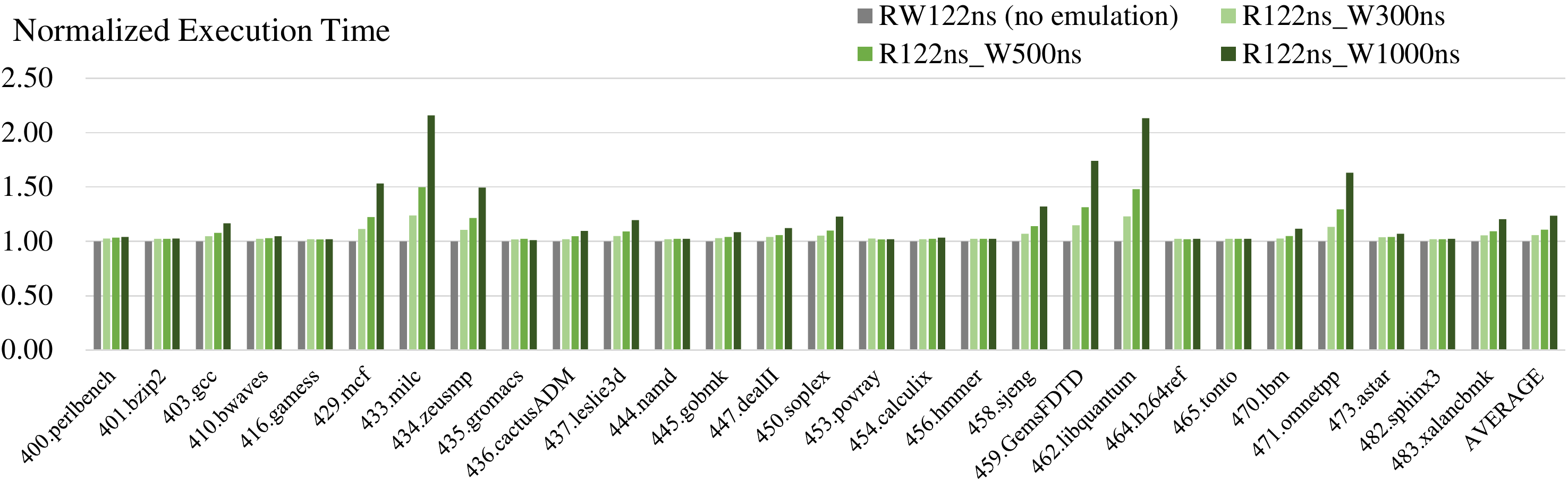}
    \end{center}
    \caption{Execution time of SPECCPU 2006 benchmark programs when emulating the read/write asymmetric latencies of NVM.
    The results are normalized to the \textit{no emulation} case.
    The target NVM write latencies were set to higher values than the DRAM latency (300 ns, 500 ns, and 1000 ns) while the target NVM read latency was always set to the same as the DRAM latency (122 ns).}
    \label{fig:allspec_time}
  \end{figure*}

  \begin{figure*}[t]
    \begin{center}
      \includegraphics[width=170mm]{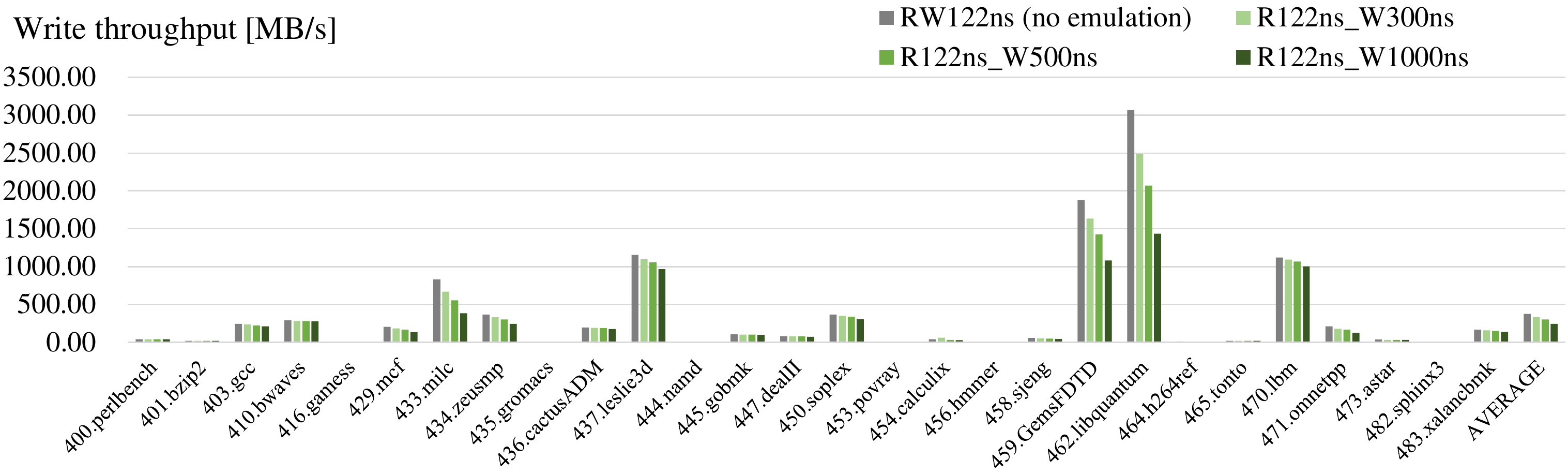}
    \end{center}
    \caption{Write throughput of each benchmark program in the emulation. The experimental condition is the same as Fig.~\ref{fig:allspec_time}.}
    \label{fig:allspec_write}
  \end{figure*}

\subsection{Applying to Various Workloads}
\label{sec:eval:speccpu2006}
To show the effectiveness of our emulation model for estimating the performance of future NVM devices, we evaluated the performance of various workloads when emulating NVM-based main memory.
We executed benchmark programs of SPECCPU 2006 and applied our prototype to them to emulate their behavior in NVM-based main memory.
We measured the execution time of each benchmark program in the emulation.
We used 28 benchmark programs mixing compute-intensive and memory-intensive workloads in the experiment.
We also measured memory write throughput of each benchmark program to see the intensity of write memory accesses.
We used an internal performance counter of the memory controller to measure write throughput.
The counter measures the total bytes written in the memory modules.
The average write throughput was calculated by dividing the total written data size by the total execution time of the benchmark.

Fig.~\ref{fig:allspec_time} shows the execution time of each benchmark program in the latency emulator.
Fig.~\ref{fig:allspec_write} shows their average write throughput.
In the experiments, we set the target NVM read latency to the same value as the DRAM read latency, while we set the target NVM write latency to 300 ns, 500 ns, and 1000 ns.
According to the results, compute-intensive workloads such as 416.gamess, 435.gromacs, and 444.namd keep their performance the same as DRAM-based main memory because they cause a small number of write-backs.
On the other hand, write-intensive workloads such as 433.milc, 459.GemsFDTD and 462.libquantum lead to the increase of the execution time and the degradation of the write throughput due to the high NVM write latency.
These results indicate that our model can emulate the behavior of practical workloads running with NVM-based main memory according to their memory access characteristics.

  \begin{figure}[t]
    \begin{center}
      \includegraphics[width=84mm]{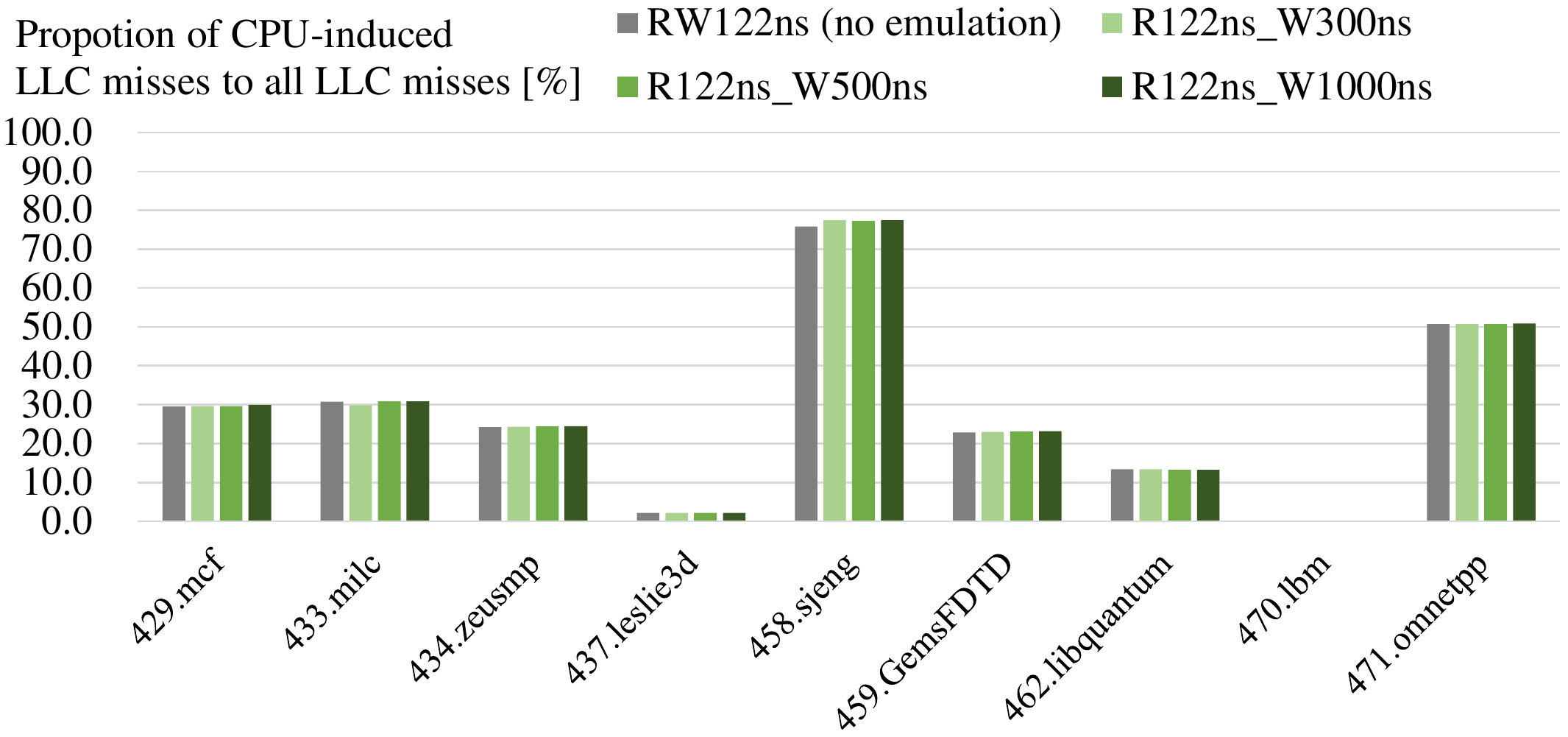}
    \end{center}
    \caption{Proportions of LLC misses induced by CPU cores to all the LLC misses during the experiments.
    }
    \label{fig:spec_llc_proportion}
  \end{figure}

  \begin{figure*}[t]
    \begin{center}
      \includegraphics[width=170mm]{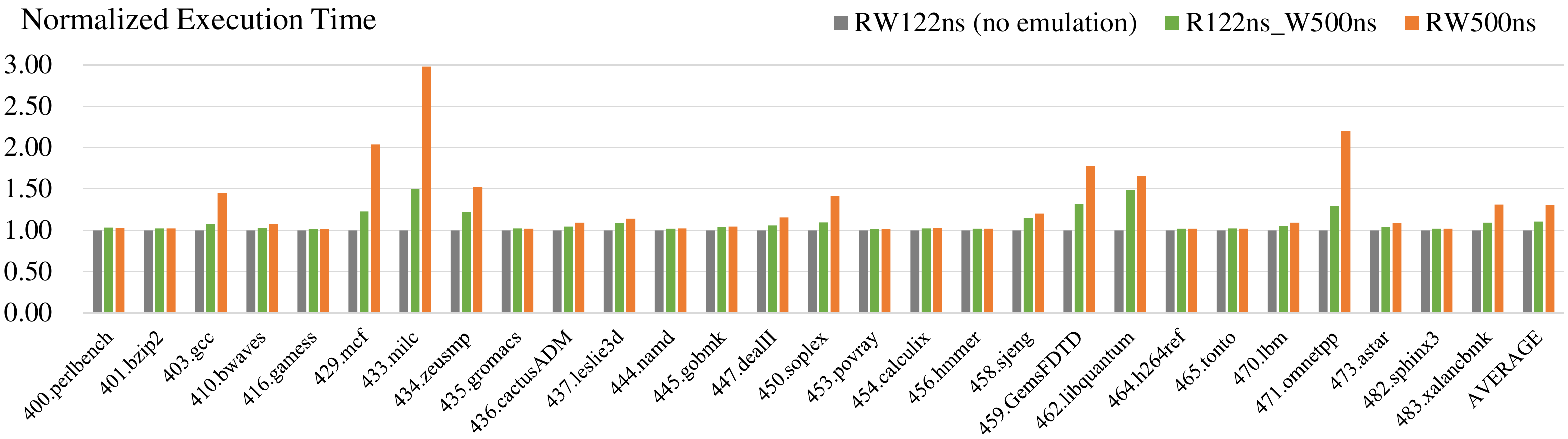}
    \end{center}
    \caption{Execution time of SPECCPU 2006 benchmarks when setting the emulated NVM read/write latencies to 500 ns.
    The results are normalized to \textit{no emulation}.
    }
    \label{fig:allspec_time2}
  \end{figure*}

Some workloads in Fig. \ref{fig:allspec_time} and \ref{fig:allspec_write} are not sensitive to their write intensiveness.
For instance, 437.leslie3d and 470.lbm are the third and fourth most write-intensive of all the benchmarks.
However, the slow down of their execution time in the emulation was less than other write-intensive workloads.
On the other hand, 458.sjeng and 471.omnetpp are less write-intensive while their execution time more sharply increased as the higher write latency was emulated.
Thus, the intensity of memory write is not only the factor that determines how a write latency impacts on workload performance.
The effectiveness of prefetchers explains the results of these applications.
Fig. \ref{fig:spec_llc_proportion} shows the proportions of LLC misses induced by CPU cores to all the LLC misses during the experiment.
The graph shows the values of the selected five write-intensive benchmarks in addition to 437.leslie3d, 470.lbm, 458.sjeng, and 471.omnetpp.
As we described in Sec. \ref{sec:problem}, LLC misses is induced by not only CPU cores but prefetchers.
As shown in Fig. \ref{fig:spec_llc_proportion}, the proportion of CPU-induced LLC misses at the execution of 437.leslie3d and 470.lbm was quite small because of the prefetchers.
Contrary, LLC misses occurred when executing 458.sjeng and 471.omnetpp were more likely induced by CPU cores themselves. 
These results show that our emulator is effective to estimate the performance impact of memory-level parallelism on NVM systems.

Since Quartz does not distinguish read/write latencies, users possibly obtain erroneous results when using it for emulating an NVM device with asymmetric read/write latencies.
Thus, we examined how each SPECCPU benchmark program behaved differently when we do not distinguish read/write latencies.
Fig. \ref{fig:allspec_time2} shows the results when we configured both NVM read/write latencies to 500 ns in the emulation.
When setting both read/write latency to 500 ns, several benchmark programs such as 429.mcf, 433.milc, and 471.omnetpp experienced more serious performance degradation than the case of setting only the write latency to 500 ns.
Since read-only LLC misses induced by these workloads are more dominant than write-back LLC misses, setting both emulated NVM read/write latencies to 500 ns resulted in the worse performance than setting only the write latency to 500 ns.
This fact indicates that the capability in emulating read/write latencies independently is indispensable for accurate emulation of NVM devices.

\renewcommand\thefootnote{\arabic{footnote}}

To clarify the slowness of cycle-accurate simulators, we measured the elapsed time of a cycle-accurate simulation for SPECCPU 2006 benchmark programs.
We set up a cycle-accurate simulation system comprising a CPU simulator (gem5~\cite{Binkert:2011:GS:2024716.2024718}) and a memory simulator (NVMain~\cite{6296505}).
We executed the 444.namd and 462.libquantum benchmark programs of SPECCPU2006 on it.
Because the simulation system is too time-consuming, we used smaller datasets to execute the benchmark programs than those used in other experiments.
Table \ref{table:host_time} shows the evaluation results.
As shown in the table, the simulation is several thousand times slower than our emulator.
The results indicate that our emulator is more light-weight than cycle-accurate simulators.

\begin{table}[t]
  \caption{Elapsed time to perform the NVM emulation/simulation. In the emulation/simulation, the NVM write latency was configured to 300 ns.}
  \begin{center}
    \begin{tabular}{c|c|c|c} \hline \hline
     & bare execution & our emulator & NVMain\&gem5 \\ \hline \hline
    444.namd & 14.4 sec &  14.7 sec & 81210.8 sec \\ \hline
    462.libquantum & 8.8 sec & 9.1 sec & 61917.5 sec \\ \hline \hline
    \end{tabular}
  \end{center}
	\label{table:host_time}
\end{table}

\subsection{A Case Study using a Realistic Workload}
\label{sec:eval:memcached}
As a case study with a realistic workload, we applied our emulator prototype to Memcached, an in-memory key-value store database.
We also chose memaslap as a client application of Memcached.
Memaslap randomly generates get/set requests following a given set/get proportion and sends them to a Memcached server during a given time period.
We executed memaslap for one minute and measured the average throughput (operations per second).
In the experiment, a Memcached server program and our emulator were executed on the machine shown in Table \ref{table:env}.
The number of Memcached worker threads was set to one\footnote[1]{We focus this paper to the validation of the emulation accuracy for single-threaded workloads. The emulation accuracy of multi-threaded workloads are discussed in future work.}.
Besides, memaslap was executed on another machine with Intel Xeon CPU E5-2650 v4 @2.2GHz.
The number of memaslap worker threads was set to eight.
The key and value sizes of each request were set to 128 bytes and 2048 bytes, respectively\footnote[2]{We chose these parameter values so that the memcached workload would cause a sufficient number of LLC misses.}.

\begin{figure}[t]
    \begin{center}
      \includegraphics[width=84mm]{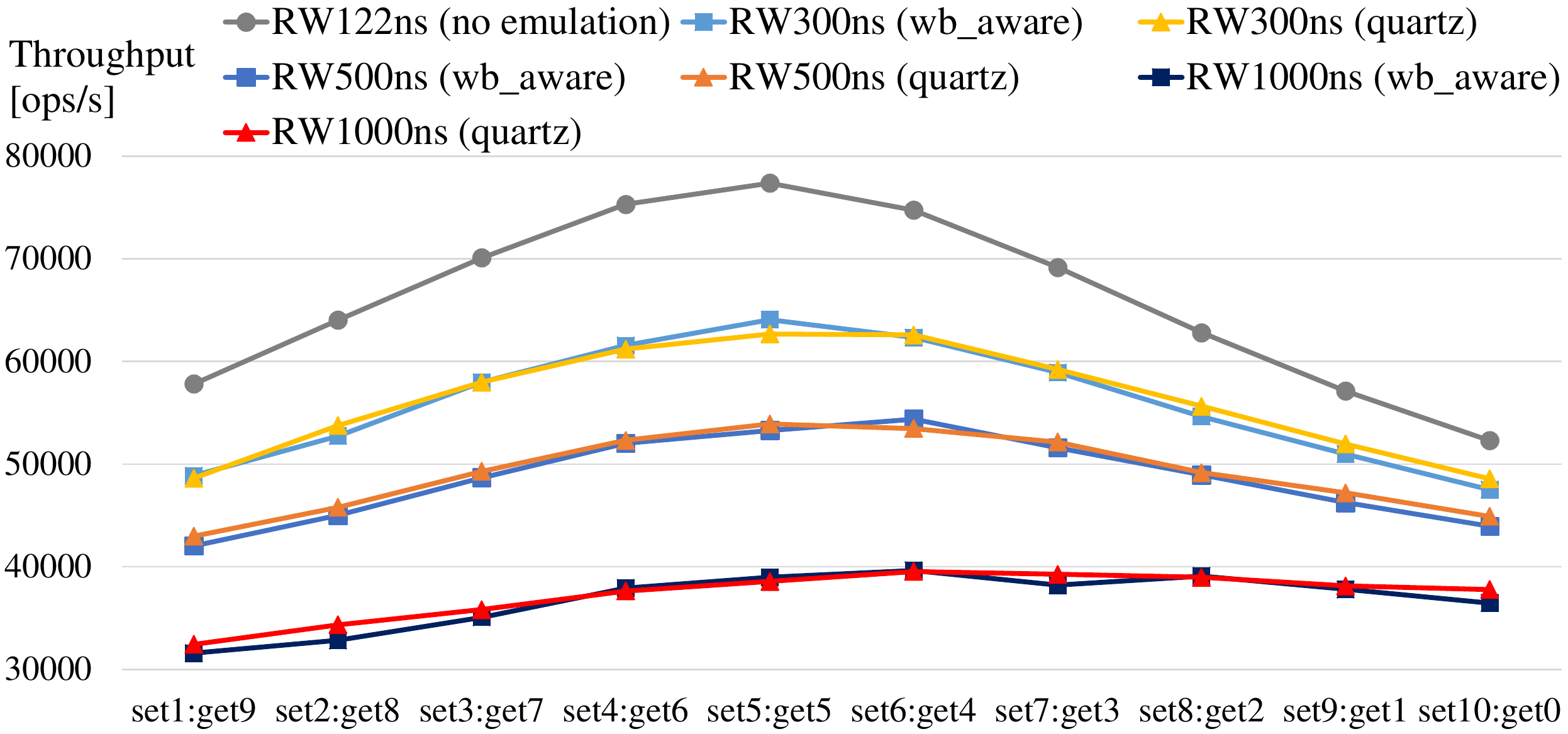}
    \end{center}
    \caption{Throughput of memaslap when setting the emulated NVM read/write latencies to the same value using two emulators: ours (wb\_aware) and an existing emulator (quartz).
    The experiments were conducted at different ratios of set/get operation.
    For example, \textit{set1:get9} means the ratio of set/get is 1:9. }
    \label{fig:memcached_samelat}
  \end{figure}

Fig. \ref{fig:memcached_samelat} shows the throughput of memaslap when setting emulated NVM read/write latencies to the same value.
We compared our model with Quartz.
In this experiment, we also executed the original Quartz on our evaluation environment and applied it to Memcached.
The evaluation results of Quartz is also shown in the figure.
It should be noted that memaslap achieves the best performance when the ratio of set:get is 5:5.
Thus, most results in the figure have peek throughput at set5:get5.
The figure shows that the throughput of memaslap decreased as the latency of NVM set higher.
When emulating the same latency, we observed nearly the same throughput in our emulator and Quartz.
This fact indicates that our emulator and Quartz are accurate to emulate main memory with symmetric read/write latencies.

\begin{figure}[t]
    \begin{center}
      \includegraphics[width=84mm]{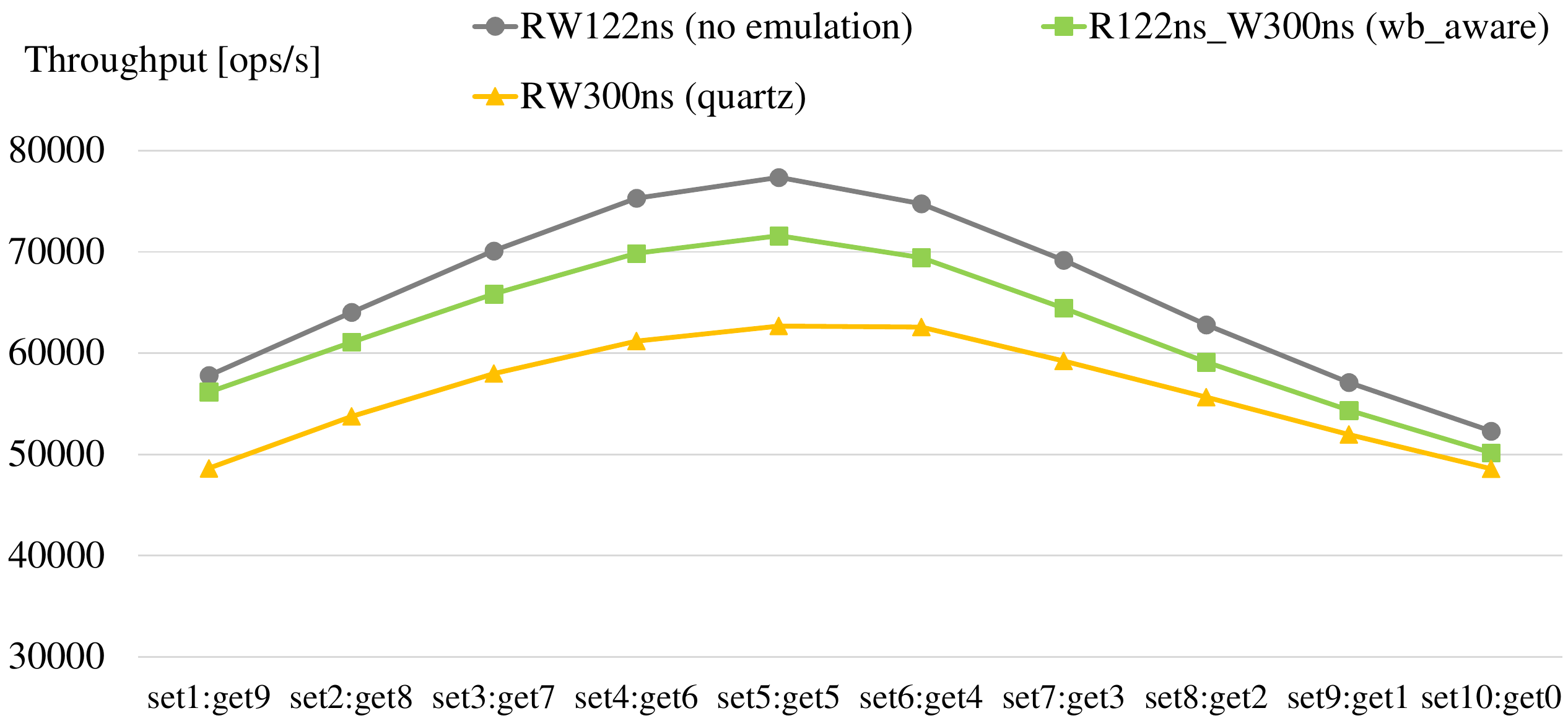}
    \end{center}
    \caption{Throughput of memaslap when setting 300 ns to the NVM write latency with our emulator (wb\_aware). The result is compared with Quartz setting 300 ns to NVM read/write latencies. }
    \label{fig:memcached_lat300ns}
  \end{figure}

\begin{figure}[t]
    \begin{center}
      \includegraphics[width=84mm]{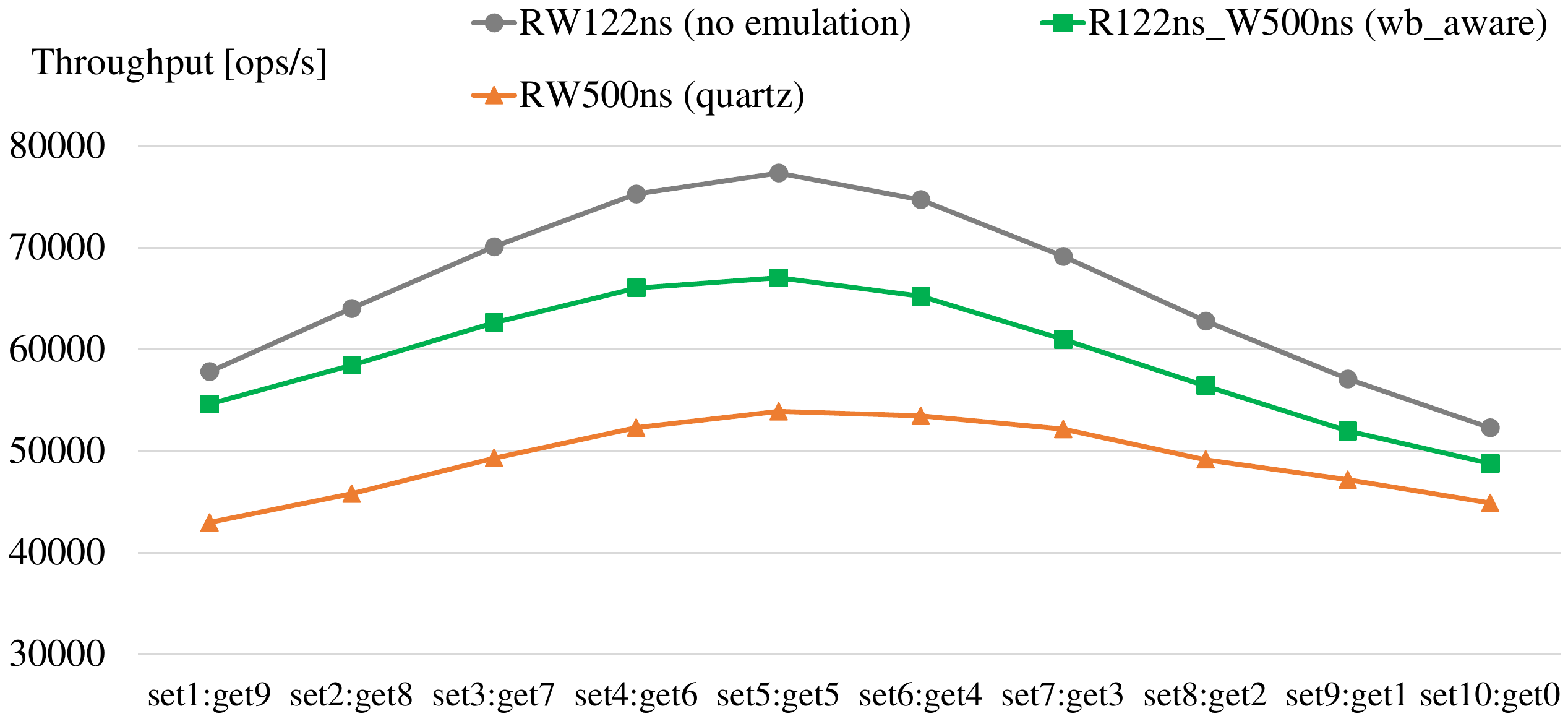}
    \end{center}
    \caption{Throughput of memaslap when setting 500 ns to the NVM write latency with our emulator (wb\_aware). The result is compared with Quartz setting 500 ns to NVM read/write latencies. }
    \label{fig:memcached_lat500ns}
  \end{figure}

\begin{figure}[t]
    \begin{center}
      \includegraphics[width=84mm]{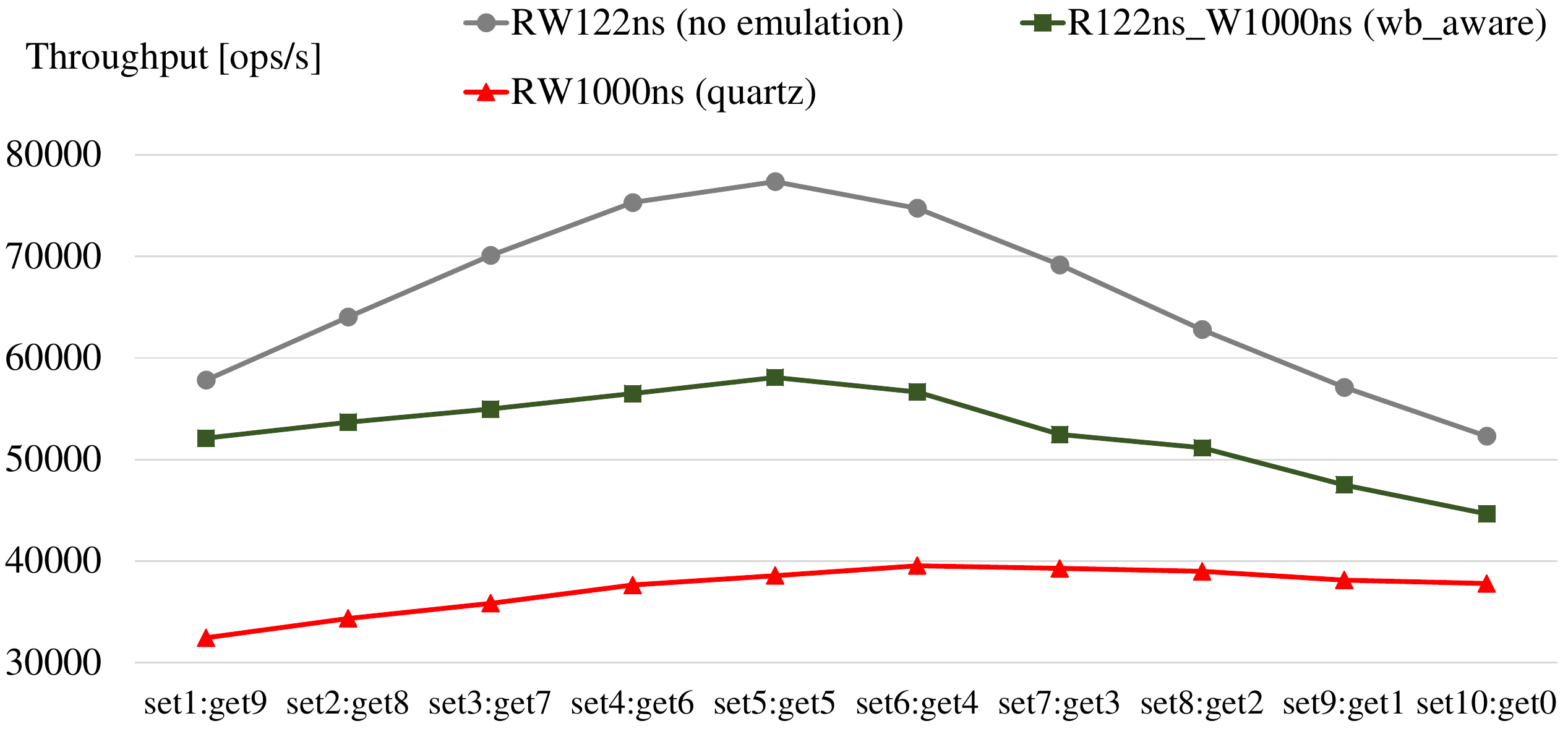}
    \end{center}
    \caption{Throughput of memaslap when setting 1000 ns to the NVM write latency with our emulator (wb\_aware). The result is compared with Quartz setting 1000 ns to NVM read/write latencies. }
    \label{fig:memcached_lat1000ns}
  \end{figure}

Fig. \ref{fig:memcached_lat300ns}, \ref{fig:memcached_lat500ns} and \ref{fig:memcached_lat1000ns} show the throughput of memaslap when the emulators were intended to emulate an NVM device with asymmetric read/write latencies.
We tried to emulate an NVM device with the read latency of 122 ns (i.e., the same as DRAM) and the write latency of 300 ns. Since Quartz does not distinguish read/write latencies, we have no choice but to set its latency to 300 ns.
As shown in the results, there is a significant performance difference between our emulator and Quartz;
in all the three figures, the throughput emulated by Quartz are lower than our emulator.
Our emulator only delays LLC misses that are expected to induce write-backs.
On the other hand, Quartz delays both read-only and write-back LLC misses since it is not aware of the difference between the two types of LLC misses.
This indicates that our emulator has great advantages in emulating read/write asymmetric memory devices. The use of Quartz likely under-estimates application performance for such NVM devices.

\section{Related Work}
Cycle-accurate simulators such as NVMain, DRAMSim2, and NVSim are widely used to evaluate software performance on NVM systems~\cite{6296505,5732229,6218223}.
In general, these memory simulators are combined with processor simulators such as Gem5 and MARSS~\cite{Binkert:2011:GS:2024716.2024718,Patel:2011:MFS:2024724.2024954}.
They calculate the full-system behavior of target architecture per CPU cycle.
This approach can set NVM read/write latencies independently while the time required for a simulation is enormous.
Our experiment found that the full-system simulation with NVMain and gem5 is approximately three orders of magnitude slower than the light-weight emulation of our proposed emulator.

Some researchers customized the hardware of commodity computer systems to accurately imitate the behavior of NVM-based main memory.
Persistent Memory Emulation Platform (PMEP) enables NVM latency emulation with special CPU microcode of an Intel Xeon processor and a customized BIOS system~\cite{Dulloor:2014:SSP:2592798.2592814}.
The microcode monitors a batch of LLC misses and injects additional delays to emulate higher NVM latency.
Lee et al. integrate an FPGA-based NVM emulator on an ARM System-on-Chip board~\cite{6966901}.
A hardware module implemented in the FPGA part monitors read/write requests issued from CPU cores to a DRAM controller and inserts additional delays to each request.
These hardware-based mechanisms can emulate slow NVM accesses with small performance overhead while such hardware customization is not easy for software researchers.

LEEF~\cite{7979841} is an NVM emulation platform that provides both full-system simulation and light-weight emulation.
The emulation mode of LEEF supports several emulation models based on existing work~\cite{Dulloor:2014:SSP:2592798.2592814,Sengupta:2015:FEN:2668930.2695529,6008552}.
However, the accuracy of these emulation models heavily depends on types of workloads; it is reported that LEEF causes emulation errors of approximately 30\% to 40\% in the worst case.
To complement the accuracy of these models, LEEF also proposes a regression method to select an optimal emulation model according to the type of a workload.
However, the detail of the regression method is not clear in this paper.

Quartz~\cite{Volos:2015:QLP:2814576.2814806} is similar work to our emulator as we described before, while Quartz is lack of support for asymmetric read/write latencies.
HME is another software-based emulator using CPU performance counters~\cite{8342227}.
HME also tries to emulate slow NVM writes by counting the number of LLC lines written back to main memory modules.
Their emulation model calculates a delay to be inserted to the execution of a target process, from the total number of write-back requests.
It, however, evenly distributes the delay to each CPU core.
This approach is not accurate because it does not consider important factors such as the per-core difference of LLC miss frequency and memory-level parallelism.  
In contrast, our emulation model can cover these factors.

\section{Conclusion}
In this paper, we presented a software-based emulation mechanism supporting asymmetric read/write latencies of NVM-based main memory.
It can emulate the behavior of NVM-based main memory, using normal DRAM-based computers.
The emulation model of our emulator inserts a delay to the execution of a target process.
It calculates the delay from the number of LLC misses and write-back operations using performance counters of the CPU cores and the LLC controller in a processor.
We implemented a prototype of the emulation model for an Intel processor family (i.e., Haswell) and evaluated its accuracy through experiments.
The results of the experiments showed that our proposed mechanism successfully emulated target read/write latencies with negligible errors of 0.1\% to 1.1\%.
We confirmed that the use of the existing emulator without the support of asymmetric latencies (i.e., Quartz) seriously under-estimated the performance of several workloads.
The use of our emulator, thanks to the modeling of the write-back mechanism of a processor, successfully generated realistic performance for these workloads.
Because emerging NVM devices such as PCM, ReRAM, and MRAM basically have asymmetric read/write latencies, our emulator has great advantages on the emulation of main memory comprising NVM.

In future work, we furthermore evaluate the accuracy of the proposed mechanism using actual NVM devices that are supposed to be available in the upcoming years.
Since the energy consumed by reading and writing NVM is different, we assume that our write-back aware emulation model is also effective for evaluating the energy performance of NVM devices.
We will clarify the effectiveness of our model for the energy asymmetry of NVM.  

\section*{Acknowledgment}
This work is supported by JSPS Grant KAKENHI 16K00115 and 19H01108.

\bibliographystyle{ieicetr}
\bibliography{ms.bib}

\profile*{Atsushi Koshiba}{%
is a Ph.D. student in the Department of Electric and Information Sciences at Tokyo University of Agriculture and Technology.
He received a master degree from the Department of Computer and Information Sciences at the same university in 2016.
His research interests include operating systems, heterogeneous computing, and energy-saving technologies for computer systems.
He is a student member of ACM, IEEE, and IPSJ.
}

\profile*{Takahiro Hirofuchi}{%
is a senior researcher of National Institute of Advanced Industrial Schience and Technology (AIST) in Japan.
He is working on system software technologies for non-volatile memory devices.
He obtained a Ph.D. of engineering in March 2007 at the Graduate School of Information Science of Nara Institute of Science and Technology (NAIST).
He obtained the BS of Geophysics at Faculty of Science in Kyoto University in Marchi 2002.
He is an expert of operating system, virtual machine, and network technologies.
}
\profile*{Ryousei Takano}{%
is a research group leader of the Institute of Advanced Industrial Science and Technology (AIST), Japan.
He received his Ph.D. from the Tokyo University of Agriculture and Technology in 2008.
He joined AXE, Inc. in 2003 and then, in 2008, moved to AIST.
His research interests include operating systems and distributed parallel computing.
He is currently exploring an operating system for heterogeneous accelerator clouds.
}
\profile*{Mitaro Namiki}{%
is a professor in the Faculty of Engineering at Tokyo University of Agriculture and Technology.
His research interests include operating systems, programing languages, parallel processing, and computer networks.
He has the Ph.D. degree in computer science from Tokyo University of Agriculture and Technology.
He is a member of ACM and IPSJ.
}

\end{document}